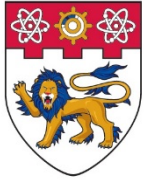

# Sustainable Development Goal (SDG) 8: New Zealand's Prospects while Yield Curve Inverts in Central Bank Digital Currency (CBDC) Era

Qionghua Chu
Nanyang Business School, Nanyang Technological University

# Sustainable Development Goal 8: New Zealand's Prospects while Yield Curve Inverts in Central Bank Digital Currency Era

By Qionghua Chu (Ruihua, Katherine), CAIA, FRM

**Abstract**

In the inverted yield curve environment, I intend to assess the feasibility of fulfilling the United Nations' Sustainable Development Goal (SDG) 8 – decent work and economic growth – by 2030 in New Zealand. Central Bank Digital Currency (CBDC) issuance supports SDG 8, based on the Cobb-Douglas production function, Solow's growth accounting relation, and Keynes's Theory of Aggregate Demand. Bright prospects exist for New Zealand.

# 1. Introduction

SDG 8, with twelve targets, of the United Nations (UN) advocates for decent work and economic growth by 2030. Intriguingly, several central banks, including the Reserve Bank of New Zealand (RBNZ), share this goal. As central banks are crucial to economic stability, I aim to evaluate RBNZ's medium- to long-term CBDC policy for SDG 8.

# 2. Current Impacts on SDG 8 in the Inverted Yield Curve Environment while CBDC Is Considered

To combat inflation, higher interest rates are necessary. Short-term rates above long-term rates invert the yield curve, which typically rises for maturity and reinvestment concerns. An inverted yield curve boosts demand for long-term government bonds like the United States (US) 10-year Treasury Notes and New Zealand (NZ) 10-year government bonds over 3-month US and NZ Treasury Bills. The former fluctuates faster in prices and yields.

First and foremost, on May 3$^{rd}$, 2023, the Federal Open Market Committee (FOMC) hiked rates to 5.00% to 5.25% for the tenth time since March 17$^{th}$, 2022 (Powell, 2023), inverting the yield curve. Despite the FOMC's halt, slower economic growth and a likely mild recession imperil the economy (Powell, 2023). After March 1985, the Impossible Trinity – free exchange rates, monetary autonomy, and an open capital account (Sullivan, 2013) – requires rate adjustments to prevent hot money from departing NZ for higher-rate nations. To manage inflation and maintain capital, especially foreign investments, the RBNZ Monetary Policy Committee (MPC) raised OCRs for the twelfth time to 5.5% on May 24$^{th}$, 2023 (RBNZ, 2023). FOMC's press releases frequently signal its future moves and hence, foreshadow RBNZ's actions (FOMC, 2023). I demonstrate in Figure 1 that the MPC cut OCRs on May 8$^{th}$, 2019 (RBNZ, 2023) based on FOMC's signals before the FOMC's October 30$^{th}$, 2019 monetary easing (Powell, 2019). Due to inflation, the FOMC lifted the FFTR on March 16$^{th}$, 2022 (Powell, 2022) and the MPC hiked rates on October 6$^{th}$, 2021 (RBNZ, 2023).

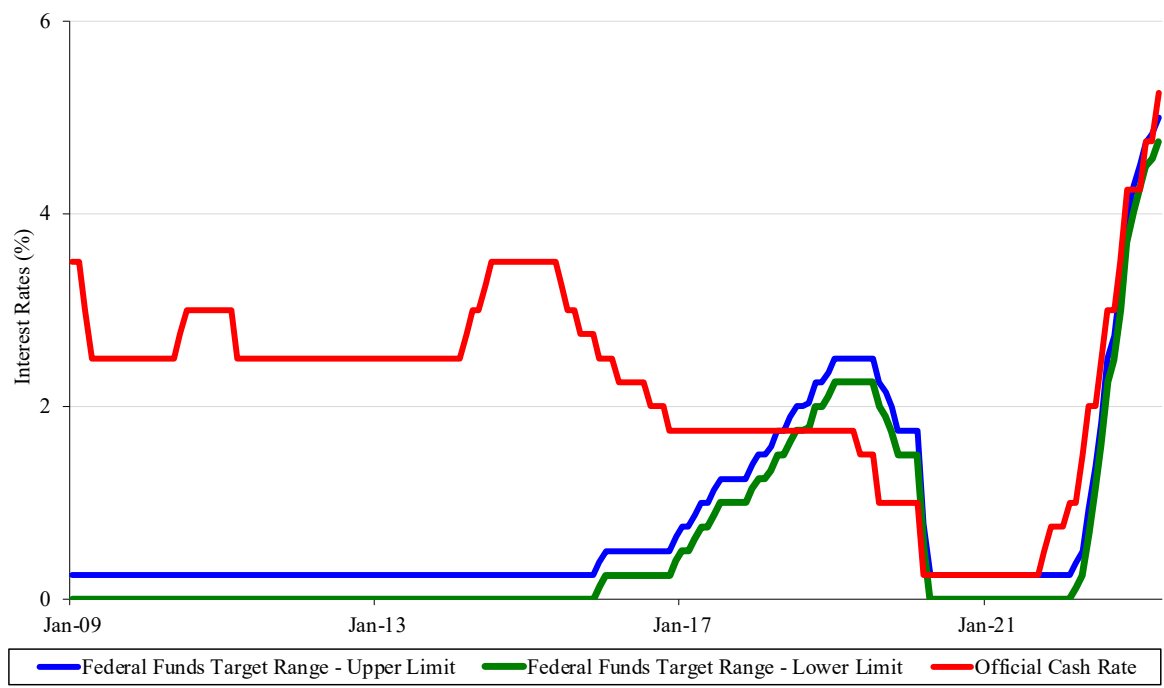

**Figure 1**. *Rates Set by the FOMC and the MPC*

Moreover, recessions commonly follow yield curve inversions and negative term spreads (Bauer & Mertens, 2018). RBNZ's dual objectives are a 2% long-term average Consumer Price Index (CPI) inflation target and maximum sustainable employment (RBNZ, 2023). Statistics NZ (SNZ) reported 6.7% year-over-year inflation in March 2023 (SNZ, 2023) and 3.4% unemployment in December 2022 (SNZ, 2023). RBNZ rate hikes may end when inflation approaches its average 1% to 3% medium-term target (RBNZ, 2023). NZ's September and December 2022 GDP grew by 1.7% and -0.6% (SNZ, n.d.). Because no two consecutive negative GDP growths exist (De Neve et al., 2018), NZ is not in recession. NZ's real GDP growth is expected to fall from 2.4% in 2022 to 1.1% in 2023 and 0.8% in 2024, then rise to 2.4% in 2025 and 2.5% in 2027 (International Monetary Fund (IMF), 2023), indicating a positive outcome.

Finally, as shown by the People's Bank of China (PBOC)'s prototype digital Renminbi, the e-CNY, CBDC may enhance technology (PBOC, 2021). NZ's CBDC consideration (RBNZ, 2021) may boost economic growth. The RBNZ's CBDC, a fiat digital currency based on political and legal confidence (RBNZ, 2021), might complement the development of Bitcoin and Ether. CBDC may make the NZ Dollar (NZD) a digital and physical legal tender, serving as a unit of account, a store of value, and a medium of exchange (RBNZ, 2023). Without account-based payments, this is quintessential. Distributed Ledger Technology stablecoins, such as Tether, bode well for digital NZD. This may aid SDG 8.

# 3. Future Impacts of CBDC Issuance on SDG 8 Realization by 2030 as Yield Curve Inverts

Under SDG 8, annual real GDP per capita and employed person growth rates measure Targets 8.1 and 8.2 of sustainable economic growth and diversity, innovation, and upgrade for economic productivity (UN, n.d.). Apart from labor (L) and capital (K) inputs, technological development, i.e., Total Factor Productivity (TFP), T, scales output, or GDP, as illustrated by Y in (1). By applying the Cobb-Douglas production function (Cobb & Douglas, 1928), (1), where α and (1 – α) denote the share of capital and labor in total factor cost and α < 1, CBDC issuance and technological advancement might boost economic growth.

$$Y = T \cdot K^{\alpha} \cdot L^{(1-\alpha)} \tag{1}$$

Additionally, dividing (1) by labor input, assuming constant capital deepening, $\frac{K}{L}$, and labor percentage in total factor cost, α, (2) shows that CBDC issuing increases labor productivity, $\frac{Y}{L}$ (Cobb & Douglas, 1928). This boosts the economy.

$$\frac{Y}{L} = T \cdot \left(\frac{K}{L}\right)^{\alpha} \tag{2}$$

However, as α is less than 1, capital deepening has a declining effect. A smaller percentage of capital in total factor cost, α, results in lower benefits of capital deepening. Since NZ is a developed nation, its higher capital-to-labor ratio and lower capital share in total factor cost make technological progress more significant. This is because T increases labor productivity more efficiently. I show in Figure 2 that technological development could boost production from L to L'. Rightward movement along L from A to A' could improve production per worker from $\frac{Y}{L}$ to $\frac{Y}{L}'$ by increasing capital deepening from $\frac{K}{L}$ to $\frac{K}{L}'$. Without technological innovation to push L upward to L' at the same $\frac{K}{L}'$, labor productivity could not have increased from $\frac{Y}{L}'$ to $\frac{Y}{L}''$ to move the economy from A' to A". Thus, CBDC issuance and software and hardware infrastructure advancements are vital for labor productivity growth.

On economic growth, (1) could be expressed as (3) – the growth accounting relation (Solow, 1957), where $\Delta \frac{Y}{Y}$, $\Delta T$, $\frac{\Delta K}{K}$, and $\frac{\Delta L}{L}$ indicate changes in GDP, technology, capital, and labor respectively.

$$\Delta \frac{Y}{Y} = \Delta T + \alpha \cdot \left(\frac{\Delta K}{K}\right) + (1 - \alpha) \cdot \left(\frac{\Delta L}{L}\right) \tag{3}$$

Moreover, (3) can be stated as (4) (Solow, 1957), with $g_{GDP}$, $g_{Long\text{-}term, technology}$, $g_{Long\text{-}term, capital}$, and $g_{Long\text{-}term, labor}$ denoting growth in GDP and long-term growth in technology, capital, and labor correspondingly. As shown in (4), CBDC could increase $g_{Long\text{-}term, technology}$ and hence, $g_{GDP}$, i.e., economic growth.

$$g_{GDP} = g_{Long\text{-}term, technology} + \alpha \cdot g_{Long\text{-}term, capital} + (1 - \alpha) \cdot g_{Long\text{-}term, labor} \tag{4}$$

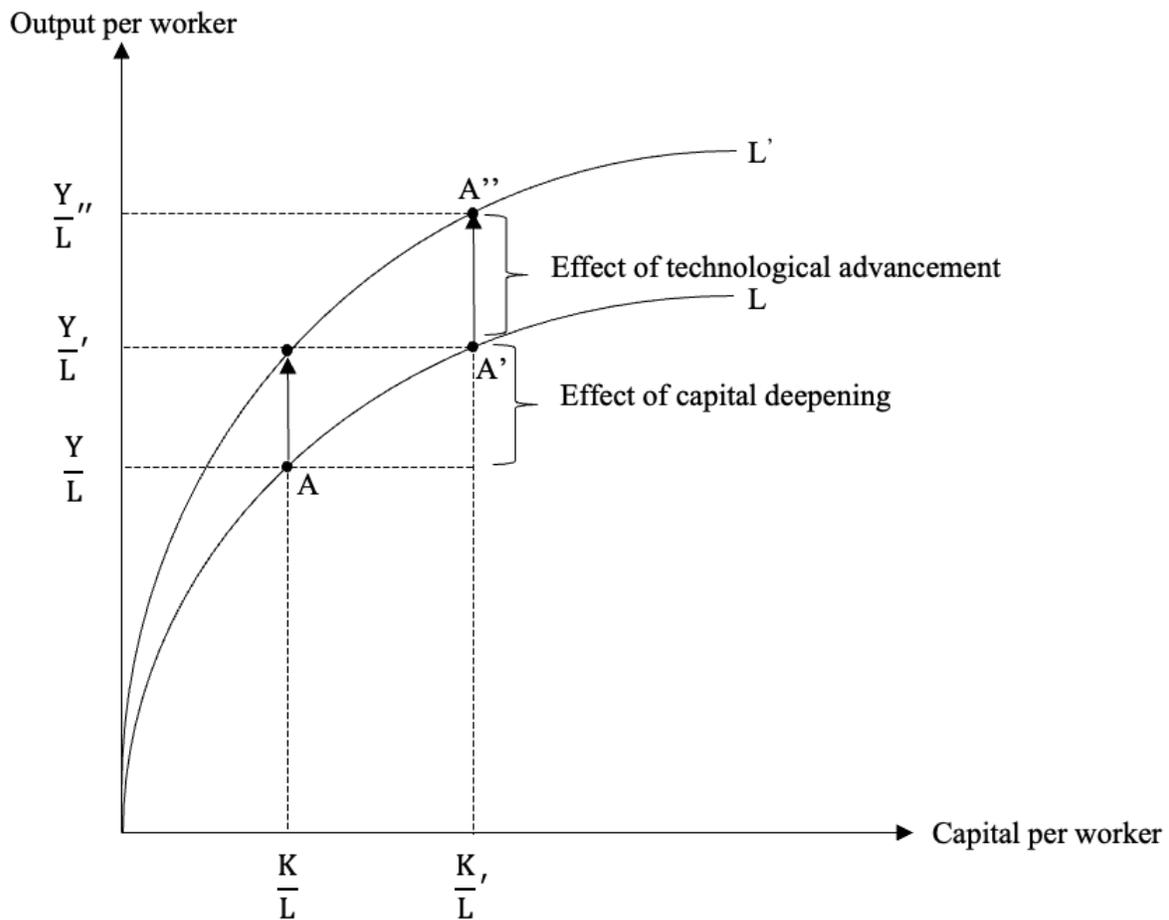

**Figure 2**. *Impact of Technological Progress on Labor Productivity*

Besides, we might form (5) (Solow, 1957) to describe (4), with $g_{\text{Long-term, labor productivity}}$ representing long-term labor productivity increase. Advancing technology, CBDC enhances $g_{\text{GDP}}$ by increasing long-term labor productivity, as seen from (5).

$$g_{\text{GDP}} = g_{\text{Long-term, labor productivity}} + g_{\text{Long-term, labor force}} \tag{5}$$

Moreover, CBDC may reduce RBNZ's monetary policy implementation time. (RBNZ, 2021). CBDC disintermediates retail and commercial banks. Because of disintermediation, RBNZ may become banks, families, and enterprises' 'lender of last resort'.

CBDC may modify asset values, exchange rates, expectations, and confidence using market rates. Exchange rates affect import prices and inflation. Aggregate demand impacts output and productivity. Output and inflation may mutually affect each other.

In addition, by the expenditure approach, according to Keynes's Theory of Aggregate Demand (1936), CBDC may simplify computerized transactions and increase consumption, C, in (6). CBDC could strengthen business confidence in the NZ economy and speed up transactions. These might elevate investment, I, and production, Y, in (6), prolonging economic prosperity.

$$Y = C + I + G + (X - M) \tag{6}$$

By fulfilling RBNZ's dual objectives of a 2% long-term inflation target and maximum sustainable employment, Targets 8.1 and 8.2 could be met.

On Target 8.3 of encouraging employment creation and enterprise development (UN, n.d.), NZ performed well. From 63.2% in 2012 to 68.1% in 2021, formal and informal employment in NZ increased (World Bank (WB), 2023). CBDC issuance may boost small- and medium-sized enterprises and commercial activities, and help to achieve Target 8.3.

Equally significant, CBDC's digital nature reduces waste and resource utilization. Abaca, linen, and other cotton paper are used to make banknotes (Cozorici et al., 2022), whereas precious metals such as copper, nickel, and zinc mint coins (Hirotsu & Chen, 2018). Also, manufacturing pollution drops. Reduced material footprint and domestic material use per capita and GDP support Target 8.4 of enhancing consumption and production resource efficiency (UN, n.d.).

Additionally, to increase high-value, less labor-intensive occupations, Target 8.5 of full employment and dignified work with fair compensation (UN, n.d.) may need modernizing hardware and software technological infrastructures. This will boost average hourly earnings and unemployment targets for all women and men, including youth and the disabled (UN, n.d.). Professional, scientific, and technology services account for 9.5% of NZ GDP (Infometrics, 2023), and boosting demand for this industry to enable CBDC issuance and wider use may create additional high-value, less labor-intensive, and equal-pay jobs. Also, the *Equal Pay Amendment Act 2020* mandates equitable pay without court orders (New Zealand Parliament, 2022). Hence, CBDC issuance helps to attain Target 8.5.

As for Target 8.6, which fosters young employment, education, and training, NZ fulfills it well by reducing the percentage of 15-to-24-year-olds not in school, work, or training from 14.25% in 2009 to 11.88% in 2021 (UN, n.d.). CBDCs may benefit NZ with positive externalities. Besides, CBDC issuance may draw young people to digital currency education, hardware and software infrastructure training, and related industry positions.

Moreover, because no 5-17-year-olds work as child laborers, NZ meets Target 8.7 to end modern slavery, trafficking, and child labor (United Nations International Children's Emergency Fund, n.d.). In April 2021, NZ, Canada, the UK, and the US questioned the International Labor Organization (ILO) on global worker rights. NZ adopted the *2014 ILO Protocol to the Forced Labor Convention, 1930* to end modern slavery in December 2019 (ILO, 2019). CBDC's low transaction reduces corporate costs, preventing forced or child labor.

NZ performs well on Target 8.8 of protecting labor rights and fostering safe working environments, as measured by fatal and non-fatal occupational injuries per 100,000 workers and national labor rights compliance (UN, n.d.). The *Employment Relations Act 2000* and the *Human Rights Act 1993* prohibit labor exploitation and protect workplace safety (Employment New Zealand, 2023). CBDC could also boost employment in less labor-intensive industries, which are less likely to cause fatal and non-fatal occupational injuries than construction, which ranks third in GDP at 6.9% in 2022 (Infometrics, 2023).

For Target 8.9, digital NZD simplifies travel and promotes positive and sustainable domestic tourism, as measured by tourist direct GDP as a share of GDP and growth rate. Domestic travelers benefit from CBDC hardware and software upgrades for cashless and cardless transactions. Faster conversion of foreign money into digital NZD in international bank and token accounts could promote international travel. CBCD issuance will increase sustainable tourism, which contributes 4.3% to NZ GDP (Infometrics, 2023), and economic growth.

For Target 8.10 of universal access to banking, insurance, and financial services, (1) the number of commercial bank branches and ATMs per 100,000 adults (over 15) and (2) the percentage of adults with bank, financial institution, or mobile-money-service provider accounts measure it. (1)

declined from 35.7 in 2009 to 17.7 in 2021 (WB, 2023), while (2) dropped from 76.38 in 2011 to 53.89 in 2021 (WB, 2023). However, rising NZ electronic card transactions bodes well. Retail card transactions rose 54.17% from NZ$5.04 billion in January 2011 (SNZ, 2011) to NZ$7.77 billion in December 2021 (SNZ, 2022). New Zealanders over 14 with bank accounts are 98.75% satisfied (Global Economy, 2023). NZ CBDC may lower (1) and raise (2). CBDC could simplify cashless transactions and increase adult bank account openings. This supports Target 8.10.

NZ achieves Target 8.a of expanding Aid for Trade (AFT) support, with total government commitments rising 682.12% from 12.86 million constant 2018 dollars in 2005 to 100.58 million in 2020 (UN, n.d.). CBDC could benefit NZ. CBDC might speed up, secure, and simplify international transactions from NZ to Least Developed Countries (LDCs), making AFT deployment easier. The NZ Ministry of Foreign Affairs and Trade (MFAT)'s Four Year Plan for LDC Tuvalu seeks strategic goals and trade (MFAT, 2021). CBDC streamlines AFT transfers to Tuvalu and boosts its economy. CBDC may advance Target 8.a.

On Target 8.b of developing an international strategy for youth employment and fulfilling the ILO's *Global Jobs Pact*, NZ's youth not in school, work, or training climbed from 10.79% in 2004 to 11.88% in 2021 (UN, n.d.). However, the upswing is transient, as 11.88% is down from 12.94% in 2020 (UN, n.d.). Since young people may find CBDC, the new money, more attractive to study and work on, CBDC issuance may promote Target 8.b.

# 4. Conclusion and Policy Considerations

In a nutshell, notwithstanding the inverted yield curve, I firmly believe that CBDC issuance in NZ could enhance economic development and provide decent jobs to fulfill SDG 8.

Since the twelve targets are interrelated, policymakers may issue CBDC to achieve some or all. When the yield curve inverts, Targets 8.1 to 8.6 can boost discretionary household income and job prospects to sustain consumption, business investment, and economic growth. In the medium to long run, yield curve normalization will boost CBDC issuance. Increased government spending on law enforcement and rights education could accomplish Targets 8.7 and 8.8, while sustainable tourism infrastructure, easier financial services, more AFT resources, and better youth employment outlook could advance Targets 8.9, 8.10, 8.a, and 8.b. Lastly, authorities can collaborate with experienced global partners to construct CBDC and achieve SDG 8.

Despite CBDC issuance risks, I envision bright prospects for SDG 8 to be realized in NZ.